\documentclass[12pt]{article}
\usepackage{epsfig}
\usepackage{latexsym}
\usepackage{color}
\newcommand{\mysquare}[0]{\raise-.2ex\hbox{{\Large$\Box$}}}
\def\lsim{\mathrel{\rlap {\raise.5ex\hbox{$ < $}}
{\lower.5ex\hbox{$\sim$}}}}
\def\gsim{\mathrel{\rlap {\raise.5ex\hbox{$ > $}}
{\lower.5ex\hbox{$\sim$}}}} \topmargin -1.5cm \textheight=22.5cm
\catcode`\@=11
\newcount\hour
\newcount\minute
\newtoks\amorpm
\hour=\time\divide\hour by60
\minute=\time{\multiply\hour by60 \global\advance\minute by-\hour}
\edef\standardtime{{\ifnum\hour<12 \global\amorpm={am}%
        \else\global\amorpm={pm}\advance\hour by-12 \fi
        \ifnum\hour=0 \hour=12 \fi
        \number\hour:\ifnum\minute<10 0\fi\number\minute\the\amorpm}}
\edef\militarytime{\number\hour:\ifnum\minute<10 0\fi\number\minute}
\def\draftlabel#1{{\@bsphack\if@filesw {\let\thepage\relax
   \xdef\@gtempa{\write\@auxout{\string
      \newlabel{#1}{{\@currentlabel}{\thepage}}}}}\@gtempa
   \if@nobreak \ifvmode\nobreak\fi\fi\fi\@esphack}
        \gdef\@eqnlabel{#1}}
\def\@eqnlabel{}
\def\@vacuum{}
\def\draftmarginnote#1{\marginpar{\raggedright\scriptsize\tt#1}}
\def\draft{\oddsidemargin -.2truein
        \def\@oddfoot{\sl preliminary draft \hfil
        \rm\thepage\hfil\sl\today\quad\militarytime}
        \let\@evenfoot\@oddfoot \overfullrule 3pt
        \let\label=\draftlabel
        \let\marginnote=\draftmarginnote
   \def\@eqnnum{(\theequation)\rlap{\kern\marginparsep\tt\@eqnlabel}%
\global\let\@eqnlabel\@vacuum}  }

\relax

\newcommand{\ba}[0]{\begin{eqnarray}}
\newcommand{\ea}[0]{\end{eqnarray}}
\def\bs{\begin{subequations}}
\def\es{\end{subequations}}

\def\thebibliography#1{%
\vskip 0.5cm \centerline{\bf References}
\list{%
[\arabic{enumi}]}{\settowidth\labelwidth{[#1]}
\leftmargin\labelwidth
\advance\leftmargin\labelsep
\usecounter{enumi}}
\def\newblock{\hskip .11em plus .33em minus .07em}
\sloppy\clubpenalty4000\widowpenalty4000
\sfcode`\.=1000\relax}

\renewcommand{\theequation}{\arabic{section}.\arabic{equation}}

\renewcommand{\section}{\setcounter{equation}{0}\@startsection%
{section}{1}{0mm}{-\baselineskip}{0.5\baselineskip}%
{\normalfont\normalsize\bfseries}}

\renewcommand{\subsection}{\@startsection%
{subsection}{2}{0mm}{-\baselineskip}{0.5\baselineskip}%
{\normalfont\normalsize\bfseries}}

\renewcommand{\subsubsection}{\@startsection%
{subsubsection}{3}{0mm}{-\baselineskip}{0.5\baselineskip}%
{\normalfont\normalsize\slshape}}
\def\crbig{\\\noalign{\vspace{3mm}}}

\usepackage{amscd} \usepackage{amsmath} \usepackage{amsfonts}

\def\Re{\,{\rm Re}\, }

\def\s{\sigma}

\def\thefootnote{\fnsymbol{footnote}}
\def\es{\end{subequations}}

\def\ec{\hat E^{c}_{2}}

\def\nn{\nonumber}

\newcommand{\uarrw}[0]{\mathrel{
{\raise.5ex\vbox{\hrule width 1cm}\hskip-6pt\rightarrow}}}

\def\bea{\begin{array}}
\def\bem{\begin{displaymath}}
\def\beq{\begin{equation}}

\def\eea{\end{array}}
\def\eem{\end{displaymath}}
\def\eeq{\end{equation}}

\def\Re{\mathop{\rm Re}}

\def\s2w{\sin^2 \theta_W}

\def\crbig{\\\noalign{\vspace {3mm}}}

\def\be{\begin{equation}}
\def\ee{\end{equation}}
\def\bc{\begin{center}}
\def\ec{\end{center}}
\def\bea{\begin{eqnarray}}
\def\eea{\end{eqnarray}}

\def\nn{\nonumber}
\begin{document}
\renewcommand{\theequation}{\arabic{section}.\arabic{equation}}
\begin{titlepage}
\begin{flushright}
NEIP--05--01 \\
LPTENS--05/01\\
CPTH--PC001.0105\\
ROMA--1400/05\\
hep-th/0503229 \\
\end{flushright}
\begin{centering}
\vspace{15pt} {\bf FLUXES AND GAUGINGS:
\boldmath
${N = 1}$ EFFECTIVE SUPERPOTENTIALS}$^\ast$\\
\vspace{15pt} {J.-P.~DERENDINGER,$^1$ C.~KOUNNAS,$^{2}$
\unboldmath
\\
P.M.~PETROPOULOS$^3$ and F.~ZWIRNER$^{4}$} \vskip .3cm {\small
$^1$ Physics Institute, Neuch\^atel University, \\
Breguet 1, CH--2000 Neuch\^atel, Switzerland \vskip .1cm $^2$
Laboratoire de Physique Th\'eorique,
Ecole Normale Sup\'erieure,$^\dagger$ \\
24 rue Lhomond, F--75231 Paris Cedex 05, France \vskip .1cm $^3$
Centre de Physique Th\'eorique, Ecole Polytechnique,$^\diamond$
\\
F--91128 Palaiseau, France \vskip .1cm $^4$ Dipartimento di
Fisica, Universit\`a di Roma `La Sapienza', and
\\ INFN, Sezione di Roma, P.le A.Moro 2, I--00185 Rome, Italy
} \vspace{5pt}

{\bf Abstract}\\
\end{centering}
We illustrate the correspondence between the $N=1$ superstring
compactifications with fluxes, the $N=4$ gauged supergravities and
the superpotential and  K\"ahler potential of the effective $N=1$
supergravity in four dimensions. In particular we derive, in the
presence of general fluxes, the effective  $N=1$  supergravity
theory associated to the type IIA orientifolds with D6 branes,
compactified on $T^6/(Z_2 \times Z_2)$. We construct explicit
examples with different features: in particular, new IIA no-scale
models, new models with cosmological interest and a model which
admits a  supersymmetric AdS$_4$ vacuum with all seven main moduli
($S, T_A, U_A,A=1,2,3$) stabilized.

\vspace{5pt} \noindent \small{\textsl{To appear in the proceedings
of the EU--RTN Workshop {\it ``The Quantum Structure of Spacetime
and the Geometric Nature of Fundamental Interactions''},
Kolymbari, Crete, September 5--10, 2004.}} \vfill
\hrule width 6.7cm \vskip.1mm{\small \small \small \noindent
$^\ast$\ Research partially supported by the EU under the
contracts HPRN-CT-2000-00131, HPRN-CT-2000-00148,
MEXT-CT-2003-509661,
MRTN-CT-2004-005104 and MRTN-CT-2004-503369.\\
$^\dagger$\ Unit{\'e} mixte  du CNRS et de l'Ecole
Normale Sup{\'e}rieure, UMR 8549.\\
$^\diamond$\  Unit{\'e} mixte  du CNRS et de  l'Ecole
Polytechnique, UMR 7644.}
\end{titlepage}
\newpage
\setcounter{footnote}{0}
\renewcommand{\thefootnote}{\arabic{footnote}}

\setlength{\baselineskip}{.7cm}
\setlength{\parskip}{.2cm}

\setcounter{section}{0}
\section{Introduction}
Superstring and M-theory compactifications admit four-dimensional
vacua with some exact or spontaneously broken supersymmetries. The
breaking scheme depends on the choice of compactification,
including orbifold or orientifold projections, and the
phenomenologically most attractive patterns are $N = 8 {\rm \ or \
} 4 \rightarrow N = 1 \rightarrow N =0$. Even with broken
supersymmetry, the underlying ten-dimensional theory encodes the
constraints of $N\geq 4$  supersymmetry, which can then be used to
derive information on the structure of the effective $N = 1$
supergravity. This effective four-dimensional theory typically
includes moduli fields originating either from the dilaton field
{$\Phi$} and the internal metric $g_{IJ}$, or from $p$-form
potentials $F_{p}$. Determining the vacuum expectation values of
the moduli fields is essential in order to (\romannumeral1) reduce
the number of massless scalars, (\romannumeral2) induce
supersymmetry breaking, (\romannumeral3) determine the
four-dimensional background geometry. This requires to generate a
potential for these scalars.

In $N\geq 4$ supergravities, the only available tool for
generating a potential is to turn abelian gauge symmetries,
naturally associated with vector fields, into non-abelian ones.
This procedure of \emph{gauging} introduces in the theory a gauge
algebra $G$ acting on the vector fields in the gravitational
and/or vector supermultiplets. The important fact is that from the
point of view of the ``daughter" $N = 1$ supergravity, the gauging
modifications only affect the superpotential $W$, whereas the
K\"ahler potential $K$ and hence the kinetic terms remain the same
as in the ungauged theory.

At the superstring level, the generation of a superpotential is in
general obtained when the background field configuration includes
non-trivial fluxes and/or Scherk--Schwarz periodicity conditions
\cite{SchSch}. Our main point is to establish the correspondence
of these data of the superstring theory, with, at the effective
field theory level, the gauging of some algebra allowed by the
massless multiplet content of the theory. It is then possible to
exhaustively study phenomenological aspects like supersymmetry
breaking, generation of masses, four-dimensional space--time
geometry, soft breaking terms, condensation phenomena,
cosmological behaviour directly in this simple field-theory
approach. This method has been developed in Ref. \cite{DKPZ} with
particular emphasis on type IIA strings, and in more general terms
in Ref. \cite{DKPZlong}. The present contribution is a summary of
these works.

\boldmath
\section{The $N = 1$ superpotential from $N=4$ gauging }
\unboldmath

To illustrate more concretely the method, consider the case of the
$N = 4$ supergravity theory corresponding to heterotic strings on
$T^6$, type II strings on orientifold or on $ K3 \times  T^2$, or
any system with 16 supercharges. For all these cases, the $N=4$
scalar manifold is \cite{N=4}
\begin{equation}
\mathcal{M} = \left( { SU(1,1) \over U (1)} \right) \times \left(
{SO(6,6 + n) \over SO (6) \times  SO (6 + n)} \right).
\end{equation}
Reducing supersymmetry to $N=1$ also truncates this scalar
manifold and the final structure depends on the specific choice of
compactification. We will focus here on a $Z^2\times Z^2$ orbifold
(or Calabi--Yau) projection, which leads to the following K\"ahler
manifold:
\begin{equation}
\mathcal{K} = \left( { SU(1,1) \over U (1)} \right)_{S}
 \times \prod_{A = 1}^{3}
\left( {SO(2,2 + n_A) \over SO (2) \times
SO(2+n_A)}\right)_{T_A,U_A,Z^I_A}.
\end{equation}
The first factor is the complex scalar field of the $N=4$
gravitational supermultiplet, which can be parameterized by the
solution
\begin{equation}
\label{Sdef1} \phi_0 -\phi_1={1\over(S+{\bar S})^{1/2}},\qquad
\qquad \phi_0 +\phi_1={S\over(S+{\bar S})^{1/2}},
\end{equation}
of the $U(1,1)$-invariant constraint
\begin{equation}
|\phi_0|^2 -|\phi_1|^2={1\over2}.
\end{equation}
The next three factors are produced by the $Z_2\times Z_2$
orbifold truncation of the vector multiplet sector. The resulting
scalars are the solutions of the $N=4$ constraints
\begin{equation}
\begin{array}{rcl}
|\sigma^1_A|^2+|\sigma^2_A|^2-|\rho^1_A|^2-|\rho^2_A|^2
-|\chi^I_A|^2 &=& \displaystyle{1\over2},\crbig
(\sigma^1_A)^2+(\sigma^2_A)^2-(\rho^1_A)^2-(\rho^2_A)^2-(\chi^I_A)^2
&=& 0.
\end{array}
\end{equation}
Specifically, we choose the solution:
\begin{equation}
\begin{array}{rclrclrcl}
\sigma^1_A &=& \displaystyle{ {{ 1+T_AU_A-(Z^I_A)^2}\over 2
Y_A^{1/2}}}, \quad &\quad \sigma^2_A &=& \displaystyle{i\, {{
T_A+U_A }\over 2 Y_A^{1/2}}}, &&& \crbig \rho^1_A &=&
\displaystyle{{{ 1-T_AU_A-(Z^I_A)^2}\over 2 Y_A^{1/2}}}, &
\rho^2_A &=& \displaystyle{i\, {{ T_A-U_A }\over 2 Y_A^{1/2}}},
\quad & \quad \chi^I_A &=& \displaystyle{{i { Z^I_A }\over
Y_A^{1/2} }}.
\end{array}
\end{equation}
This choice of parameterization of the physical fields is
appropriate for string compactifications, since it singles out the
geometric moduli and dilaton as $\Re T_A$, $\Re U_A$, $\Re S$.
Each string construction is then characterized by its own
complexification of these seven real scalars and by specific
couplings to matter fields $Z^I_A$. The index $A$ labels the three
complex planes defined by the $Z_2 \times Z_2$ symmetry used for
the orbifold projection.

Equipped with this set of physical fields, we can simply extract
the K\"ahler potential and the superpotential  of the $N=1$
theory, by direct inspection of the gravitino mass matrix of the
$N=4$ theory submitted to the $Z_2\times Z_2$ projection, which
reads \begin{equation}
\begin{array}{l}
\mathrm{e}^{K/2}W = (\phi_0 -\phi_1 ) f_{IJK} \, \Phi^I_1 \,
\Phi^J_2 \, \Phi^K_3 +(\phi_0 +\phi_1){\bar f}_{IJK}\,  \Phi^I_1\,
\Phi^J_2\, \Phi^K_3, \crbig
\hspace{1. cm} {\rm with} \hspace{1. cm} \Phi^I_A = \left\{
~\sigma^1_A,~\sigma^2_A ; ~\rho^1_A,~\rho^2_A,~\chi^I_A \right\},
\end{array}
\end{equation}
where $f_{IJK}$ and ${\bar f}_{IJK}$ are the structure constants
of the gauge algebra. They differ, however, in their dependence
with respect to the $S$-field through the $S$-duality phases. With
our solutions, the non-holomorphic part defines the K\"ahler
potential, \begin{equation} K=- \ln \left(S+{\bar S}\right) -
\sum_{A=1}^3 \ln Y_A, \end{equation} where
\begin{equation} Y_A = (T_A+{\bar T}_A)(U_A+{\bar U}_A)-
\sum_I(Z_A^I +{\bar Z}_A^I )^2. \end{equation} And the holomorphic
part defines the superpotential generated by the structure
constants $f_{IJK}$ and the $S$--duality phases.

With Eqs. (\ref{Sdef1}), the superpotential has two terms. The
first contribution involves the structure constants $f_{IJK}$ and
does not depend on $S$. The second term involves the second set of
structure constants $\bar f_{IJK}$ and is linear in
$S$.\footnote{Consistency constraints apply on the structure
constants: $S$-duality phases should be compatible with the gauge
algebra, and, of course, Jacobi identities should be verified.} In
the heterotic construction where the geometric moduli are $\Re
T_A$ and $\Re U_A$, these two contributions would respectively
correspond to a perturbative ``electric gauging" and a
non-perturbative ``magnetic gauging".

\section{Gauging and fluxes}

A gauging with non-zero $f_{IJK}$ generates a superpotential and
then, in particular, a potential for the moduli. Furthermore, if
the structure constants are such that the fields
$$
\sigma^1_A,~\sigma^2_A;~ \rho^1_A,~\rho^2_A
$$
are involved, the superpotential also induces supersymmetry
breaking. These directions correspond to the $N = 4$ graviphotons
and the non-abelian gauge algebra also involves the $N=4$
$R$-symmetry.

In string and M-theory, the non-trivial {$~f_{IJK}~$} and {$~{\bar
f}_{IJK}$} are generated by non-zero electric and/or magnetic
fluxes, R--R and fundamental $p$-form fields:
\begin{itemize}
\item
three-form fluxes $H_3$, in the heterotic string and in the NS--NS
sector of type IIA and type IIB theories;
\item
fluxes of $p$--forms $F_p$ in M-theory and in the R--R sector of
type IIA and type IIB;
\item
fluxes of gauge two-forms $\tilde H_2$, in heterotic ($E_8 \times
E_8$ or $SO (32)$) as well as in type I, and spin-connection
fluxes $\omega_3$ in all strings and M-theory.
\end{itemize}
Many particular cases have already been considered in the
literature by direct study of the ten- or eleven-dimensional field
equations, including specific choices of fluxes. In particular:
\begin{itemize}
\item
$H_3$ fluxes in heterotic superstrings  \cite{h3het}, including
spin connection and Yang--Mills fluxes~\cite{kalmy};
\item
fluxes of $\omega_3$, $H_3$ and $\tilde H_2$, as generated in
exact string solution via freely-acting orbifold (the
generalization to superstring theory of the Scherk--Schwarz
gauging) and group manifold compactifications \cite{SchSch2};
\item
IIB strings with simultaneous NS--NS ($H_3$) and R--R ($F_3$)
fluxes \cite{h3IIB}.
\end{itemize}
The fact that such configurations can also be studied in the
gauging approach of the effective field theory does not
necessarily shed new light on some of these cases.

The simplicity of our approach using gauging of the underlying
$N=4$ supersymmetry algebra allows however to study exhaustively
more complex cases, including also spin-connection geometric
fluxes. This was the main motivation of Ref. \cite{DKPZ}, where we
concentrated on orientifolds of type IIA strings which offered the
broadest structure of allowed fluxes and had been explored to a
lesser extent \cite{h3IIA}. The results of this study is reported
in the next two sections.

\section{Fluxes in type IIA orientifold on  \boldmath{$ T^6/ Z^2\times Z^2$}}

The $Z_2\times Z_2$ orbifold projection used in this paper is only
compatible with the orientifold admitting $D6$--branes. The
possible fluxes in this type IIA orientifold are as follows
\cite{h3IIA, aft}. Firstly, in the R--R sector, the IIA theory
compactified to four dimensions generates all even-form fluxes,
$$
F_0, ~ F_2, ~ F_4, ~ F_6
$$
(where the subscript gives the degree of the form in the compact
directions only). The mass parameter of massive IIA supergravity
and the ten-dimensional R--R four-form in space--time only, $\sim
\lambda\epsilon_{\mu\nu\rho\sigma}$, induce then $F_0$ and $F_6$,
while internal values of the R--R two- and four-forms generate
$F_2$ and $F_4$. The NS--NS sector provides then three-form fluxes
$H_3$,  as well as the geometrical fluxes arising from background
values of spin connections $\omega_3$.

To identify the superpotential terms related to the above fluxes,
we firstly need to find the appropriate complexification of the
seven moduli fields $$ S,~ T_1,~ T_2,~ T_3,~ U_1,~ U_2,~ U_3
$$
compatible with $N = 1$ supersymmetry. It is actually simpler to
begin with a discussion of the heterotic case, and then to find
the non-trivial change of variables relating heterotic and IIA
strings.  For heterotic strings on ${T^6/ Z^2\times Z^2}$, the
dilaton $\Phi$ is singled out as the gauge coupling, geometric
moduli have a natural definition in terms of the internal metric
$G_{IJ}$ and complexification involves the metric and the two-form
field: \begin{equation} \label{geom}
\begin{array}{c}
\left( G_{IJ} \right)_A = \displaystyle{\frac{t_A}{u_A}} \left(
\begin{array}{cc}
u_A^2 +\nu_A^2 & \nu_A
\\
\nu_A  & 1 \end{array} \right), \crbig T_A=t_A+i\left(
B_{IJ}\right)_A, \qquad U_A= u_A+i\nu_A ,\crbig \displaystyle{
\mathrm{e}^{-2\Phi}=s(t_1t_2t_3)^{-1}, \qquad S=s+ ia, \qquad
g_{\mu\nu}=s^{-1} {\tilde g}_{\mu\nu} }.
\end{array}
\end{equation} These heterotic definitions of moduli $s, t_A, u_A$ are also
called {\it geometrical variables} herebelow.

The supersymmetric complexification in the type IIA orientifold
with D6-branes is more involved, due to the dilaton rescaling, the
presence of R--R fields and also the orientifold projection which
in particular eliminates the geometrical modes $\nu_A$ of the
metric. Since the NS--NS two-form field $B_2$ is odd in this
orientifold, it is clear that $T_B$ moduli in IIA and heterotic
are the same: \begin{equation} T_{B, \mathrm{\, IIA}}=T_B
\qquad\qquad (B=1,2,3). \end{equation} Inspection of the
complexification of scalar kinetic terms and also of D6-brane
gauge kinetic terms indicates that the redefinitions
\begin{equation} \label{redef1} s_{\mathrm{IIA}} = \sqrt{s\over
u_1u_2u_3}, \quad u_{1,\mathrm{\, IIA}} = \sqrt{su_2u_3\over u_1},
\quad u_{2, \mathrm{\, IIA}} = \sqrt{su_1u_3\over u_2},  \quad
u_{3,\mathrm{\, IIA}} = \sqrt{su_1u_2\over u_3} \end{equation} of
the real moduli scalars are required so that the $N=1$ complex
scalars receive their imaginary part from the four components of
the three-form R--R field which survive the orbifold projection:
\begin{equation} \label{redef2}
\begin{array}{l}
S_{\mathrm{IIA}} = s_{\mathrm{IIA}} + i A_{6810}, \crbig
U_{1,\mathrm{\, IIA}} = u_{1,\mathrm{\, IIA}} + iA_{679}, \quad
U_{2,\mathrm{\, IIA}} = u_{2,\mathrm{\, IIA}} + i A_{589}, \quad
U_{3,\mathrm{\, IIA}} = u_{3,\mathrm{\, IIA}} + i A_{5710}.
\end{array}
\end{equation} In the next section, we will omit the subscript ``{IIA}", but
it should be understood that the fields defined by Eqs.
(\ref{redef1}) and (\ref{redef2}) will be used whenever we discuss
superpotential contributions.

\section{ \boldmath{$N=1$} superpotentials versus IIA fluxes}

With the definitions (\ref{redef1}), (\ref{redef2}) and Eqs.
(\ref{geom}), it is now easy to translate the effect of IIA fluxes
into specific contributions to the $N=1$ superpotential. In this
section, we merely enumerate the various contributions and
describe some examples with combined fluxes. More detail can be
found in Refs. \cite{DKPZ} and \cite{DKPZlong}. Compact dimensions
will be labelled $5,6,\ldots,10$, with $5,7,9$ odd and $6,8,10$
even under the orientifold $Z_2$.

\subsection{Single fluxes and their superpotentials}

The R--R sector generates four types of fluxes: $F_0$ and $F_6$
arise from switching on the four-form field in space--time
(Freund--Rubin ansatz) and from the mass parameter of massive IIA
supergravity \cite{romans}; switching on the two- and four-form
fields in internal directions generates $F_2$ and $F_4$.
\begin{itemize}
\item {\bf  \boldmath{$F_6$} flux}

Suppose that we switch on an internal R--R six-form $F_6$. Using
geometrical variables as defined in Eqs. (\ref{geom}), a scalar
potential of the form $V=F_6^2 / (s^2t_1t_2t_3)$ in these
variables is generated. Changing to IIA variables using
(\ref{redef1}) and (\ref{redef2}) and comparing with the $N=1$
supergravity potential leads to the superpotential
\begin{equation} W = F_6. \end{equation}

\item{\bf \boldmath{$F_0$} flux}

Similarly, switching on an internal (real) zero-form $F_0$ leads
to the superpotential \begin{equation} W = -i F_0\, T_1T_2T_3
\end{equation} in IIA variables. Via complexification, this
superpotential introduces terms depending on imaginary parts of
$T_A$, {\it i.e.} terms depending on the NS--NS two-form field
$B_2$. These terms are as predicted by the equations of massive
IIA supergravity with mass parameter $F_0$.

\item{\bf \boldmath{$F_2$} fluxes}

The orbifold and orientifold projections allow internal two-form
fluxes $F_{56}$, $F_{78}$ and $F_{910}$. The induced
superpotential is \begin{equation} W = -F_{56}\, T_2 T_3 -
F_{78}\, T_3T_1 - F_{910}\, T_1 T_2. \end{equation}

\item{\bf \boldmath{$F_4$} fluxes}

Finally, the orbifold and orientifold projections allow internal
four-form fluxes $F_{5678}$, $F_{78910}$ and $F_{56910}$. The
induced superpotential is \begin{equation} W = iF_{5678}\, T_3 +
iF_{78910}\, T_1 + iF_{56910}\, T_2. \end{equation}
\end{itemize}
In the NS--NS sector, we need to consider $H_3$ fluxes and
geometric fluxes of the spin connection $\omega_3$.

\begin{itemize}
\item{\bf \boldmath{$H_3$} fluxes}

The directions in $H_3$ allowed by the orbifold and orientifold
projections are $H_{579}$, $H_{5810}$, $H_{6710}$ and $H_{689}$.
These fluxes generate the superpotential \begin{equation} W  =
iH_{579}\, S + iH_{5810}\, U_1+ iH_{6710}\, U_2+iH_{689}\, U_3 .
\end{equation}

\item{\bf  Geometric \boldmath{$\omega_3$} fluxes}

The projections allow background values of several components of
the spin connections. These fluxes produce two categories of
superpotential terms. Firstly, \begin{equation} W =  -\omega_{679}
\,S T_1 - \omega_{895}\, ST_2 - \omega_{1057}\, ST_3.
\end{equation} Both $H_3$ and $\omega_3$ are then sources for the
$S$--dependent, ``non-perturbative" part of the superpotential.
Secondly, geometric fluxes allow to generate all bilinar
contributions of the form $T_A U_A $ or $-T_AU_B$ ($B\ne A$). For
instance,
$$
W = \omega_{6810}\,T_1U_1 + \omega_{8106} \, T_2U_2+
\omega_{1068}\, T_3U_3
$$
will be used in the next paragraph.

\end{itemize}

\subsection{Type IIA  examples with combined fluxes, gauging
and moduli stabilization}

The set of allowed fluxes in this IIA orientifold is rich enough
to provide examples of combined fluxes where some or all moduli
are stabilized. We give here only a brief description of some of
these cases.

\vspace{3mm}\noindent {\bf $\bullet$ Flat gaugings, no-scale
models, stabilization of four moduli}

\noindent If four moduli are stabilized, the resulting scalar
potential is positive definite with the flat directions of a
no-scale model \cite{noscale}.

\begin{enumerate}
\item {\it
Standard perturbative Scherk--Schwarz superpotential from
$\omega_3$ fluxes}

The superpotential reads: \begin{equation} W = a \,(T_1U_1 +
T_2U_2). \end{equation} The scalar potential is semi-positive,
$V\geq 0$. At the minimum, the four imaginary parts of $T_1, T_2,
U_1$ and $U_2$ vanish and one quadratic condition applies on the
real parts: $t_1u_1 = t_2u_2 \equiv v^2$. Then, at the minimum,
$W= 2av^2$ and the gravitino mass reads:
$$
m_{3/2}^2={|a|^2 \over 32st_3u_3},
$$
which depends on the flat directions $S, T_3, U_3$, but not on the
three flat directions left at the minimum in directions
$t_1,t_2,u_1$ and $u_2$. This model is a gauging of the
two-dimensional euclidean group $E_2$.

\item {\it A ``non-pertubative" no-scale example with $\omega_3, F_2,H_3$ and $F_6$
fluxes}

The superpotential of this example is \begin{equation} W = a (ST_1
+ T_2T_3) +ib( S+T_1T_2T_3). \end{equation} The potential is again
semi-positive, the gravitino mass is
$$
m_{3/2}^2={a^2 + b^2 \over 32 u_1u_2u_3},
$$
and there are in addition two ``decoupled" flat directions.

\item
{\it A no-scale model based on a gauging of the euclidean group
$E_3 \times E_3$}

This gauging is equivalent to non-zero fluxes of $\omega_3$,
$F_0$, $F_2$ and $H_3$. Specific choices of these fluxes allow the
following superpotential: \begin{equation} W=a(ST_1+ST_2+ST_3)+
a(T_1T_2+T_2T_3+T_3T_1) + 3ib(S+T_1T_2T_3). \end{equation} Four
moduli are stabilized, and the resulting no-scale model has
gravitino mass given by
$$
m_{3/2}^2={9\over32} {a^2 + b^2 \over u_1u_2u_3}.
$$

\end{enumerate}

\vspace{3mm}\noindent {\bf $\bullet$ Gaugings with positive
definite potential}

\noindent Examples can be easily found, in which less than four
moduli are stabilized and the potential is always strictly
positive-definite, leading to runaway solutions (in time).

Superpotentials with a single monomial are of course examples
where no modulus gets stabilized. For instance, we can choose the
fluxes $F_6$, $F_0$ or $H_3$, leading to
\begin{equation}
W=F_6 \, , \qquad W= -iF_0 \, T_1 T_2 T_3 \qquad {\rm or} \qquad
W=i H_3 S \, .
\end{equation}
This leads to $V=4 \, m_{3/2}^2 > 0$ and the gravitino mass term
is of the form
\begin{equation}
m_{3/2}^2  = {1 \over 2^7 \, s \, t_1 \, t_2 \, t_3 \, u_1 \, u_2
\, u_3} \, \times \Bigl\{ |F_6|^2 \,\, , \,\, \, |F_0 T_1 T_2
T_3|^2 \,\, {\rm or} \,\, |H_3S|^2 \Bigr\} \, ,
\end{equation}
respectively.

An example where three moduli are stabilized is obtained by
switching on a system of R--R fluxes $(F_0,F_2,F_4,F_6)$, with
superpotential
\begin{equation}
W = A \left(1 + T_1 T_2 +T_2 T_3 + T_3 T_1\right) + i
B\left(T_1+T_2+T_3+T_1T_2T_3\right) \, .
\end{equation}
This choice of fluxes and superpotential is actually a gauging of
$SO(1,3)$. It is immediate to see that, since the superpotential
does not depend on four of the seven main moduli (the $T$-moduli
are stabilized at one), supersymmetry is broken and a
positive-definite runaway $D=4$ scalar potential is generated,
\begin{equation}
V = m^2_{3/2}  \, , \qquad {\rm with} \qquad m^2_{3/2} = {A^2+B^2
\over 8 \, s \, u_1 \, u_2 \, u_3} \, ,
\end{equation}
possibly leading to time-dependent vacua of cosmological interest.

\vspace{3mm}\noindent {\bf $\bullet$ Gaugings with
negative-definite potential }

\noindent Situations where more than four moduli are stabilized
lead to negative-definite potentials once the stabilized moduli
are set to their vacuum values.

We begin with a gauging of $E_3$ with fluxes $\omega_3$
(geometric) and $F_6$ (R--R six-form). The R--R six-form
corresponds to the $SO(3)$ directions in $E_3$ while $\omega_3$
corresponds to the translations. The superpotential reads
\begin{equation}
W = \omega_3\left(T_1 U_1 + T_2 U_2 + T_3 U_3\right) - F_6 \, .
\end{equation}
The six equations for the non-trivial supergravity auxiliary
fields are solved at $\tau_A =  \nu_A  =0$ and $ t_1 \, u_1 = t_2
\, u_2 = t_3\, u_3  = F_6 / \omega_3$. At these values, $W = 2 \,
F_6$, and the $s$--dependent scalar potential and gravitino mass
term read:
\begin{equation}
V   = - 2 m_{3/2} ^2 = - {\omega_3^3 \over 16\, F_6 \, s} .
\end{equation}
At the string level, this is the well-known NS five-brane solution
plus linear dilaton \cite{KPR90AFK}, in the near-horizon limit.
The original gauging is $SU(2)$, combined with translations, which
emerge as free actions at the level of the world-sheet conformal
field theory. It is remarkable that this $E_3$ algebra remains
visible at the supergravity level. It is also interesting that, if
we allow extra fluxes, induced by the presence of
fundamental-string sources, we can reach AdS$_3$ background
solutions with stabilization of the dilaton. All moduli are
therefore stabilized. This has been studied recently at the string
level \cite{IKP}.

\vspace{3mm}\noindent {\bf $\bullet$ Stabilization of all moduli}

\noindent Using all fluxes admissible in IIA, $Z_2\times Z_2$
strings, we can obtain the stabilization of all moduli in AdS$_4$
space--time geometry. Switching on all fluxes ($\omega_3, H_3,
F_0, F_2, F_4, F_6$), we can obtain the superpotential
\begin{eqnarray}
 W & = & A \left[ 2 \ S \ (T_1+T_2+T_3) - (T_1 T_2 + T_2 T_3 +
T_3T _1 )
+ 6 \ (T_1 U_1 +T_2 U_2 + T_3 U_3 ) - 9 \right] \nn \\
& & + \,  i \ B \left[ 2 \ S + 5 \ T_1 T_2 T_3 + 2 \ (U_1 + U_2 +
U_3 ) - 3 \ ( T_1 + T_2 + T_3) \right] . \label{2ads4}
\end{eqnarray}
This is a consistent $N=4$ gauging only if
\begin{equation} 6 \, A^2 = 10 \, B^2  , \label{cond} \end{equation}
as a consequence of Jacobi identities, as shown in Refs.
\cite{DKPZ, DKPZlong}. Notice that this condition relates the even
and odd terms in the superpotential, thus its sign ambiguity is
irrelevant. The superpotential (\ref{2ads4}) leads to a
supersymmetric vacuum at $S =  T_A  = U_A = 1$ ($A=1,2,3$). Since
at this point $ W  = 4 \, ( 3 \, A + i \, B) \ne 0$, implying $V =
- 3 \, m_{3/2}^2 <0$, this vacuum has a stable AdS$_4$ geometry
with all seven main moduli frozen.

Notice that the appearance of non-integer flux numbers is actually
an artifact of our choice for presenting the model, with
$S=T_A=U_A=1$ at the minimum. One can recover integer flux numbers
by rescaling appropriately the moduli. A possible choice (among
many others) is the following:
\be (S,T_A,U_A) \; \to \; b \, (S,T_A,U_A) \, , \qquad b =
\frac{B}{A} = \sqrt{3 \over 5} \, . \ee
With that choice
\bea W & = & N \left[ 2\, S\left(T_1+T_2+T_3\right) -
 \left(T_1T_2 + T_2T_3 + T_3T_1\right)
 \right.  \nn \\
& & + \left.
    6\left(T_1U_1 +T_2U_2 + T_3 U_3 \right) -15 \right]  \nn \\
& & + i N \left[ 2S + 3 T_1T_2T_3 + 2 \left(U_1+U_2+U_3\right)
-3\left(T_1+T_2+T_3\right) \right] \, , \eea
where $N = (3/5) \, A$.

\section{Conclusion}

We should firstly emphasize that our last example is the
\emph{only} known case with complete stabilization of the moduli,
reached in IIA by switching on fundamental fluxes. We should also
stress that this cannot happen in the heterotic string, because of
the absence of $S$-dependence in the general flux-induced
superpotential. Such a dependence could appear with gaugino
condensation. In type IIB with D3-branes (and D7), the orientifold
projection that accompanies the $Z_2 \times Z_2$ orbifold
projection eliminates the $\omega_3$ fluxes, thus the $T$ moduli
are not present in the superpotential and cannot be stabilized by
fluxes. The case of D9-branes (open string) is similar to the
heterotic case.

Our approach of supergravity gauging can be viewed as a bottom-up
approach to the problem of generating moduli and matter
superpotentials in superstring vacuum configurations. As in all
approaches essentially based on effective Lagrangians, it is not
in principle expected that low-energy symmetries are powerful
enough to completely replace a top-bottom analysis using
ten-dimensional string or M-theory equations. Our studies of many
examples \cite{DKPZ, DKPZlong}, in heterotic and type II strings,
actually shows that the effective supergravity approach based upon
$N=4$ gaugings accurately reproduces the conditions imposed by the
full field equations of the ten-dimensional theories. This of
course requires to include all necessary brane and orientifold
plane contributions to these equations.

\section*{Acknowledgments}
We would like to thank the organizers of various meetings where
some of the authors had the opportunity to present the above
results: the EU--RTN Workshop {\it ``The Quantum Structure of
Spacetime and the Geometric Nature of Fundamental Interactions''},
Kolymbari, Crete, September 5--10, 2004, the workshop {\it
``Frontiers Beyond the Standard Model II''}, Minnesota, October
14--17, 2004, the {\it ``String Vacuum Workshop''},
Max--Planck--Institut f\"ur Physik und Astrophysik, Munich,
November 22--24, 2004 and the Inaugural Meeting of the EU--MRTN
programme {\it ``The Quest for Unification''}, CERN, December
6--8, 2004. This work was supported in part by the EU under the
contracts HPRN-CT-2000-00131, HPRN-CT-2000-00148,
MEXT-CT-2003-509661, MRTN-CT-2004-005104 and MRTN-CT-2004-503369.


\newpage
\begin{thebibliography}{99}
%

\bibitem{SchSch}
J.~Scherk and J.H.~Schwarz, Phys.\ Lett.\ B {\bf 82} (1979) 60 and
Nucl.\ Phys.\ B {\bf 153} (1979) 61;

\bibitem{DKPZ}
J.-P. Derendinger, C. Kounnas, P.M. Petropoulos and F. Zwirner,
arXiv:hep-th/0411276 (2004).

\bibitem{DKPZlong}
J.-P. Derendinger, C. Kounnas, P.M. Petropoulos and F. Zwirner, to
appear.

\bibitem{N=4}
A.~Das, Phys.\ Rev.\ D {\bf 15} (1977) 2805;
E.~Cremmer and J.~Scherk, Nucl.\ Phys.\ B {\bf 127} (1977) 259;
E.~Cremmer, J.~Scherk and S.~Ferrara, Phys.\ Lett.\ B {\bf 74}
(1978) 61:
%
A.H.~Chamseddine, Nucl. \ Phys.\ B {\bf 185} (1981) 403:
%
J.-P.~Derendinger and S.~Ferrara, {\it Lectures given at Spring
School of Supergravity and Supersymmetry, Trieste, Italy, Apr
4-14, 1984}, CERN-TH-3903;
%
M.~de Roo, Nucl. \ Phys. \ B {\bf 255} (1985) 515 and
Phys.\ Lett.\ B {\bf 156} (1985) 331;
E.~Bergshoeff, I.G.~Koh and E.~Sezgin, Phys.\ Lett.\ B {\bf 155}
(1985) 71;
M.~de Roo and P.~Wagemans, Nucl.\ Phys.\ B {\bf 262} (1985) 644
and
Phys.\ Lett.\ B {\bf 177} (1986) 352;
P.~Wagemans, Phys.\ Lett.\ B {\bf 206} (1988) 241;
P.~Wagemans, {\it Aspects of N=4 supergravity}, Ph.D. Thesis,
Groningen University report RX-1299 (1990).

%
\bibitem{h3het}
J.-P.~Derendinger, L.E.~Ibanez and H.-P.~Nilles, Phys.\ Lett.\ B
{\bf 155} (1985) 65 and
Nucl.\ Phys.\ B {\bf 267} (1986) 365;
M.~Dine, R.~Rohm, N.~Seiberg and E.~Witten, Phys.\ Lett.\ B {\bf
156} (1985) 55;
A.~Strominger, Nucl.\ Phys.\ B {\bf 274} (1986) 253;
R.~Rohm and E.~Witten, Annals Phys.\  {\bf 170} (1986) 454.

%
\bibitem{kalmy}
N.~Kaloper and R.C.~Myers, JHEP {\bf 9905} (1999) 010
[arXiv:hep-th/9901045].

%
\bibitem{SchSch2}
R.~Rohm, Nucl.\ Phys.\ B {\bf 237} (1984) 553;
C.~Kounnas and M.~Porrati, Nucl.\ Phys.\ B {\bf 310} (1988) 355;
S.~Ferrara, C.~Kounnas, M.~Porrati and F.~Zwirner, Nucl.\ Phys.\ B
{\bf 318} (1989) 75;
M.~Porrati and F.~Zwirner, Nucl.\ Phys.\ B {\bf 326} (1989) 162;
C.~Kounnas and B.~Rostand, Nucl.\ Phys.\ B {\bf 341} (1990) 641;
I.~Antoniadis, Phys.\ Lett.\ B {\bf 246} (1990) 377;
I.~Antoniadis and C.~Kounnas, Phys.\ Lett.\ B {\bf 261} (1991)
369.
E.~Kiritsis and C.~Kounnas, Nucl.\ Phys.\ B {\bf 503} (1997) 117
[arXiv:hep-th/9703059];
E.~Kiritsis, C.~Kounnas, P.M.~Petropoulos and J.~Rizos, Nucl.\
Phys.\ B {\bf 540} (1999) 87 [arXiv:hep-th/9807067];
I.~Antoniadis, E.~Dudas and A.~Sagnotti, Nucl.\ Phys.\ B {\bf 544}
(1999) 469 [arXiv:hep-th/9807011] and
Phys.\ Lett.\ B {\bf 464} (1999) 38 [arXiv:hep-th/9908023];





I.~Antoniadis, J.-P.~Derendinger and C.~Kounnas, Nucl.\ Phys.\ B
{\bf 551} (1999) 41 [arXiv:hep-th/9902032].

%
\bibitem{h3IIB}
J.~Michelson, Nucl.\ Phys.\ B {\bf 495} (1997) 127
[arXiv:hep-th/9610151];
K.~Dasgupta, G.~Rajesh and S.~Sethi, JHEP {\bf 9908} (1999) 023
[arXiv:hep-th/9908088];
T.R.~Taylor and C.~Vafa, Phys.\ Lett.\ B {\bf 474} (2000) 130
[arXiv:hep-th/9912152];
P.~Mayr, Nucl.\ Phys.\ B {\bf 593} (2001) 99
[arXiv:hep-th/0003198];
G.~Curio, A.~Klemm, D.~Luest and S.~Theisen, Nucl.\ Phys.\ B {\bf
609} (2001) 3 [arXiv:hep-th/0012213];
S.B.~Giddings, S.~Kachru and J.~Polchinski, Phys.\ Rev.\ D {\bf
66} (2002) 106006 [arXiv:hep-th/0105097];
S.~Kachru, M.B.~Schulz and S.~Trivedi, JHEP {\bf 0310} (2003) 007
[arXiv:hep-th/0201028];
A.R.~Frey and J.~Polchinski, Phys.\ Rev.\ D {\bf 65} (2002) 126009
[arXiv:hep-th/0201029].

\bibitem{h3IIA}
J.~Polchinski and A.~Strominger, Phys.\ Lett.\ B {\bf 388} (1996)
736 [arXiv:hep-th/9510227];
I.~Antoniadis, E.~Gava, K.S.~Narain and T.R.~Taylor, Nucl.\ Phys.\
B {\bf 511} (1998) 611 [arXiv:hep-th/9708075];
S.~Gukov, C.~Vafa and E.~Witten, Nucl.\ Phys.\ B {\bf 584} (2002)
69 [Erratum-ibid.\ B {\bf 608} (2001) 477] [arXiv:hep-th/9906070];
S.~Gukov, Nucl.\ Phys.\ B {\bf 574} (2000) 169
[arXiv:hep-th/9911011].

\bibitem{aft}
C.~Angelantonj, S.~Ferrara and M.~Trigiante, JHEP {\bf 0310}
(2003) 015 [arXiv:hep-th/0306185] and
Phys.\ Lett.\ B {\bf 582} (2004) 263 [arXiv:hep-th/0310136].

\bibitem{romans}
L.J.~Romans, Phys.\ Lett.\ B {\bf 169} (1986) 374.

%
\bibitem{noscale}
E.~Cremmer, S.~Ferrara, C.~Kounnas and D.V.~Nanopoulos, Phys.\
Lett.\ B {\bf 133} (1983) 61.

\bibitem{KPR90AFK}C. Kounnas, M. Porrati and B. Rostand, Phys.\
Lett.\ B {\bf 261} (1991) 369; I. Antoniadis, S. Ferrara and C.
Kounnas, Nucl.\ Phys.\ B {\bf 421} (1994) 343
[arXiv:hep-th/940207].

\bibitem{IKP}
D.~Isra\"el, C.~Kounnas and P.M.~Petropoulos, JHEP {\bf 0310}
(2003) 028 [arXiv:hep-th/0306053].

\end{thebibliography}
\end{document}